\documentclass[12pt]{iopart}

\usepackage{iopams, graphicx, wrapfig}

\begin{document}

\title{Strong Gravitational Lensing by Sgr A*}

\author{Amitai Y. Bin-Nun}
\address{Department of Physics and Astronomy, University of Pennsylvania, 209 S. 33$^{\rm rd}$ St., Philadelphia, Pennsylvania, USA}
\ead{binnun@sas.upenn.edu}

\begin{abstract}
In recent years, there has been increasing recognition of the potential of the galactic center as a probe of general relativity in the strong field. There is almost certainly a black hole at Sgr A* in the galactic center, and this would allow us the opportunity to probe dynamics near the exterior of the black hole. In the last decade, there has been research into extreme gravitational lensing in the galactic center. Unlike in most applications of gravitational lensing, where the bending angle is of the order of several arc seconds, very large bending angles are possible for light that closely approaches a black hole. Photons may even loop multiple times around a black hole before reaching the observer. There have been many proposals to use light's close approach to the black hole as a probe of the black hole metric. Of particular interest is the property of light lensed by the S stars orbiting in the galactic center. This paper will review some of the  attempts made to study extreme lensing as well as extend the analysis of lensing by S stars. In particular, we are interested in the effect of a Reissner-Nordstrom like $1/r^2$ term in the metric and how this would affect the properties of relativistic images.

\end{abstract}

\vspace{2pc}

\maketitle

\newpage
\section{Introduction}

\begin{figure}[h]
 \begin{center}
 \includegraphics[width=.7 \textwidth]{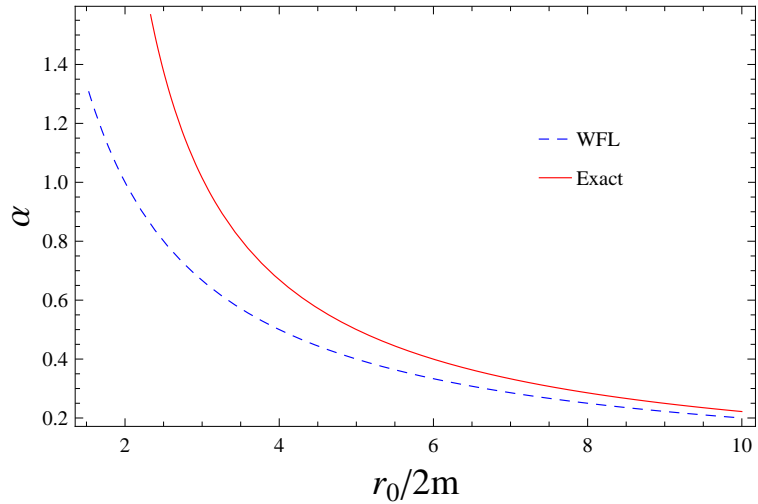}
 \caption{\label{fig:weakexact}  The weak field approximation and exact bending angle as a function of closest approach. The $x$ axis is in terms of Schwarzschild radii. This figure shows that, except for lensing in the immediate vicinity of very compact objects, the weak field approximation is close to exact.}
 \end{center}
 \end{figure}

Gravitational lensing continues to be a major source of insight into gravitation and cosmology \cite{LensingReview}. Excitingly, increasingly precise observations of the compact radio source Sgr A* at the Galactic center and its surrounding stars have given us very high confidence that only a very gross deviation from GR could allow for the absence of a black hole there \cite{propSgrA}.  The black hole is estimated to have a mass of about $4.31 \times 10^6 M_{\odot}$ and a distance of about 8.33 kpc from Earth. Black holes are unique laboratories for gravitational lensing because their compactness allows light to closely approach the photon sphere and its path will bend significantly there due to gravity. The nature of lensing is sensitive to the metric and matter composition at the very center of the galaxy. Most studies of lensing use the weak field limit approximation, but when a photon closely approaches a  black hole's photon sphere, this approximation does not hold (see Fig. \ref{fig:weakexact}).

For a spherically symmetric, static metric with line element
\begin{equation}
ds^2= -A(r)dt^2 + B(r) dr^2 +C(r) r^2 d\Omega^2,
\label{metric}
\end{equation}
the bending angle is an elliptic integral based on the functions of the metric \cite{VE2002} and is
\begin{eqnarray}
\nonumber \alpha (r_0) &=& 2 {\int_{r_0}}^{\infty}\left(\frac{B(r)}{C(r)}\right)^{1/2}
                       \left[(\frac{r}{r_0})^2\frac{C(r)}{C(r_0)}\frac{A(r_0)}{A(r)}-1\right]^{-1/2} \\ 
& \times & \frac{dr}{r}- \pi.
      \label{bending}
\end{eqnarray}
As $r_0$ gets smaller, the bending angle becomes larger, diverging at the photon sphere at $r_{\rm m}$, where photons are in a circular, unstable orbit.

Despite this phenomenon being known for a long time \cite{darwin}, the potential for studying lensing effects due to a large bending angle are only recently being realized. In this article, we will discuss the potential for using lensing in the galactic center as a probe of modified gravity, particularly looking for deviations from GR that take the form of a $1/r^2$ term in the metric. The most generic form of this metric  is the ``tidal Reissner-Nordstrom" (TRN) metric \cite{dmptidalrn}, which is 
\begin{equation}
ds^2 = -(1-\frac{2M}{r} + \frac{Q}{r^2})dt^2 + (1-\frac{2M}{r}+  \frac{Q}{r^2})^{-1} dr^2 +r^2 d\Omega^2,
\label{eq:trnmetric}
\end{equation}
with $ Q= q\:4M^2$ and $q$ being a free parameter. The TRN metric differs from the Reissner-Nordstrom metric in that $Q$ is allowed to be both positive and negative. While this metric should be seen as a generic modification of graivty, it is intriguing because the $1/r^2$ correction term is suggestive of higher dimensional theories, and the TRN metric is a possible solution for the black hole hole on the brane in the Randall-Sundrum II model. Background on the RS braneworlds, black holes in RS braneworlds, and the applicability of the TRN metric for use in a large black hole can be found in \cite{whiskerthesis, me, me2, methesis}.

In Sec. \ref{sec:rel}, we will talk about the recent devlopment of relativistic images and their theoretical use in probes to distinguish a Schwarzschild black hole from a black hole in modified gravity, particular a braneworld black hole. As can be seen from the form of the bending angle, the exact properties of gravitational lensing are dependent on the metric functions. An excellent review of mathematical formalism for lensing by black holes is in \cite{bozzareview}. This paper will concentrate more on the astrophysical predictions made by recent studies of black hole lensing as well show some new results. In Sec. \ref{sec:sgra}, we will discuss the more recent application of lensing to the secondary images of S stars orbiting in the galactic center as well as present new results regarding the relativistic images of such images. We conclude in Sec. \ref{sec:conc}.

\section{Relativistic Images}
\label{sec:rel}

\begin{figure}[ht!]
\begin{minipage}[b]{0.35 \textwidth}
\centering
\includegraphics[scale=1]{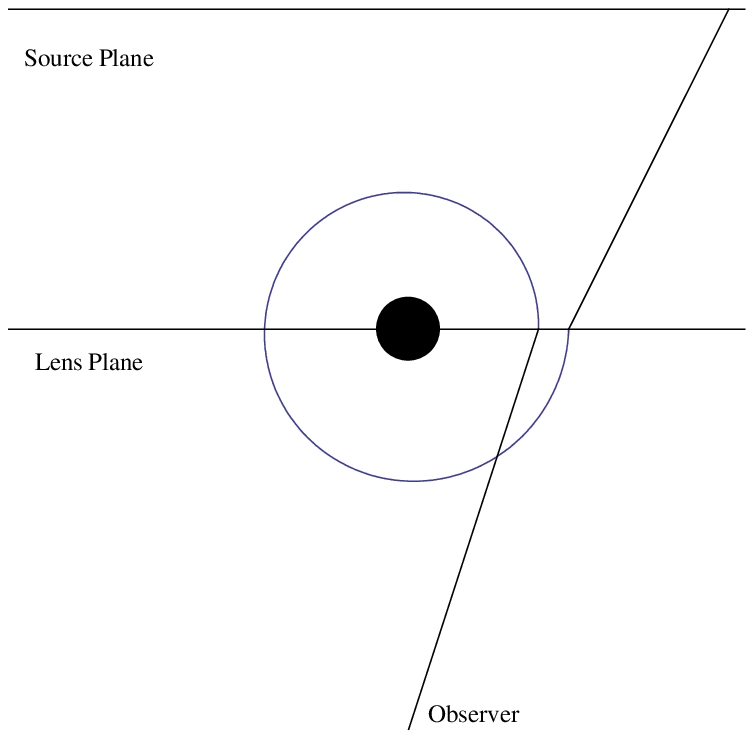}
\end{minipage}
\hspace{0.5cm}
\begin{minipage}[b]{0.35 \textwidth}
\centering
\includegraphics[scale=1]{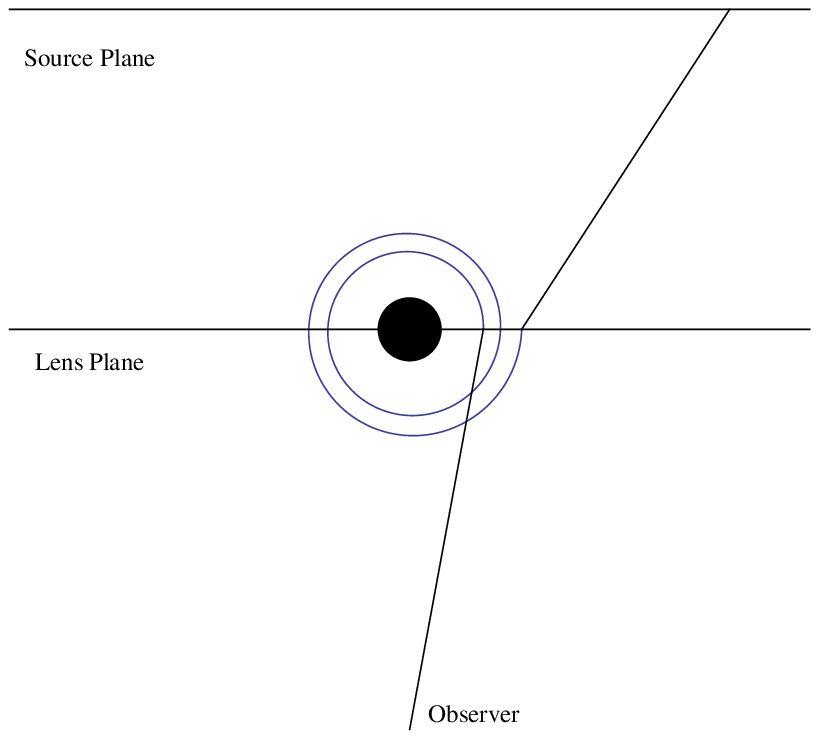}
\end{minipage}
 \caption{\label{fig:loop} A sketch of the null geodesics forming the first two relativistic images on the primary image side of the optic axis. The figure on the left represents the first relativistic image, which forms when a photon loops around the black hole once. The figure on the right represents the second relativistic image which forms even closer to the black hole, after looping around twice. In reality, both images are very close together on the lens plane and very close to the black hole's photon sphere.}
\end{figure}

In 2000, Virbhadra and Ellis \cite{VE2000} revived interest in the study of very strong gravitational lensing. Because black holes are extremely compact, they are able to cause photons to loop around the black hole once or more times before they reach the observer. This is schematically illustrated in Fig. \ref{fig:loop}. They calculated the location and magnitude of the outermost two relativistic images on both sides of the source for a variety of image positions. They showed that for all realistic image positions, relativistic images are extremely demagnified. In later work, \cite{VE2002, VK2008} showed that relativistic image properties near Sgr A* will be different if the Schwarzschild metric is substituted for a Janis-Newman-Winicour metric associated with a static, spherically symmetric real scalar field and catalogued the dependency of relativistic image properties on scalar charge. This demonstrated the potential for extreme lensing to differentiate between different types of matter in the galactic center. The work done in these papers was done by numerically solving the Virbhadra-Ellis lens equation for the image position $\theta$ from the optic axis and for the image magnification. Details can be found in \cite{VE2000, me, methesis}.

Early attempts at developing an analytic approach to relativistic images were made by \cite{darwin}. More recently, \cite{bozzaetal} expressed Eq. (\ref{bending}) as an expansion of an elliptic integral which diverged at the photon sphere. They found the first order expansion of the elliptic equation from its divergence at the photon sphere and used it in place of Eq. (\ref{bending}). This approach was updated in \cite{bozza2002, bozza2007} to include arbitrary source and observer positions. This approach is later used to calculate lensing observables in the braneworld scenario \cite{eiroad}. In \cite{bozza2002, me}, analysis showed that for large bending angles ($\alpha > \pi$), the analytic approach and numerical approach are nearly identical except in very special spacetimes. 

Later studies showed the results of lensing of light around Sgr A* if the mass at Sgr A* was assumed to be a Reissner-Nordstrom black hole \cite{eiroarn}, a braneworld black hole \cite{eiroad, whisker2005, casadio, mmreview, aliev, me}, or other exotic spacetimes \cite{exotic1, exotic2, exotic3}. Work on braneworld black holes found that a negative value for $Q$ in the TRN metric creates a larger photon sphere, so relativistic images are further from the optic axis than in a Schwarzschild spacetime. In addition, they find that using the TRN metrics causes the relativistic images to cluster closer to the black hole --- there is not as much of a separation between the first two relativistic images as there is in a Schwarzschild spacetime. Finally, having a negative value of $Q$ makes the ratio between the brightness of first image and the sum of the rest of the images (or just the secondary) image larger. This means that the first image is brighter compared to the second image in the TRN spacetime.  In \cite{me}, we studied the lensing effects of primordial black holes which had fixed quantities of tidal charge.

The upcoming Multi-AO Imaging Camera for Deep Observations (MICADO) telescope at the European Extremely Large Telescope (E-ELT) \cite{micado} is projected to come online in about 2018 and have a maximum sensitivity of about the $30^{\rm th}$ magnitude in the near infrared and resolution in the astrometric mode of about $10 \: \mu$ arc sec. While the astrometric resolution is close to the scale on which relativistic images operate, MICADO will not be nearly sensitive enough to observe relativistic images. In addition, the galactic center is not a clean environment for observations. Clearly, relativistic images will not be part of observational astronomy for the foreseeable future. If the observation of relativistic images is nearly impossible, using the properties of relativistic images to differentiate between theories will be even more difficult and is even further in the future. 

\section{Using S Stars as Sources}
\label{sec:sgra}

While it may not be feasible to observe relativistic images, there are have been other lines of inquiry into lensing where the bending angle is large and observational prospects are not as dismal. Holz and Wheeler \cite{holzwheeler} consider the idea that a small black hole of roughly a solar mass can lens the light from the sun and redirect it back to Earth. In the case when the Earth is in between the Sun and the black hole and on the line connecting the two, the bending angle is $\pi$. They termed such a black hole a ``retro-MACHO" and they find that a $10 M_{\odot}$ black hole at $10^{-2}$ pc from the sun (and on the opposite side of the Earth) and nearly perfectly aligned will produce a ring with a magnitude of $26.1$. However, this value drops off sharply as the black hole falls out of maximum alignment.  In \cite{depaolis}, the effect of ``retro-lensing", or lensing with a bending angle of about $\pi$, is studied for the case of the star S2, which orbits around the galactic center, finding a maximum brightness in the $K$ band of $m_K =30$ before extinction of light from the dust in the galactic center is considered. Finally, by using orbital parameters of the S stars in the galactic center, \cite{bozzas2} study the properties of secondary images of those S stars. We will examine this possibility more closely.

Many stars orbiting close to the black hole at Sgr A* have been observed carefully and their orbits can be reconstructed from orbital parameters published in \cite{propSgrA}. At each point in time, the star is treated as a source being lensed by the black hole. Using orbital parameters and the intrinsic brightness of the stars in the K-band, \cite{bozzas2} calculated the position and magnitude of the secondary and relativistic images of many stars, assuming a Schwarzschild metric. Over time, the images would get brighter and fainter, as the star would get more and less aligned, respectively, with the optic axis. In \cite{me2}, we extended the analysis of secondary images of S stars to the TRN metric. We showed that if the value of $Q$ in Eq. \ref{eq:trnmetric} is negative, the brightness of the secondary image for all times is increased. For a large enough value of $Q$, this increase in brightness can be significant. In addition, we considered positive values of $Q$ including an extremal Reissner-Nordstrom (ERN) black hole. We showed that, for positive $Q$, the secondary images will be fainter, but there will not be a very significant difference between the brightness of the image in the Schwarzschild spacetime and in the extremal Reissner-Nordstrom spacetime. For further elaboration, see \cite{me2}.

In this paper, we will extend this analysis to relativistic images, and will examine the first relativistic image of the star S2. Image positions are obtained by solving a modification of the Ohanian lens equation \cite{me2, bozzalens}:
\begin{equation}
\gamma= \alpha(\theta)-\frac{D_L}{D_{LS}}\theta,
\label{eq:ohanian}
\end{equation}
where $\gamma$ is the angle between the optic axis and the line connecting the source to the lens, $\theta$ is the image position (to the observer), $D_L$ is the constant distance from the observer to the lens (in this case, the distance between us and Sgr A*) and $D_{LS}$ is the distance, which varies over time, between the lens and the source star. In \cite{bozzas2, me2}, the Keplerian orbits are solved, yielding the functions $\gamma(t)$ and $D_{LS}(t)$. We are particularly interested in the behavior of the star S2 and its orbitial parameters are in Table \ref{table:stars}. The magnification of an image at any given point is given by
\begin{equation}
\mu= \frac{D_L^2}{D_{LS}^2}\frac{\sin \theta}{\frac{ d\gamma}{d \theta} \sin \theta}.
\label{eq:mag}
\end{equation}
\begin{table*}[t]
\caption{\label{table:stars}Orbital Parameters of S2:  $a$ is the semimajor axis, $e$ is the eccentricity, $i$ is the inclination of the normal of the Orbit with respect to the line of sight, $\Omega$ is the position angle fo the ascending node, $\omega$ is the periapse anomaly with respect to the ascending node, $t_P$ is the epoch of either the last or next periapse, $T$ is the orbital period, and $K$ is the apparent magnitude in the $K$ band (date taken from \cite{propSgrA})
\newline}
\scalebox{.77}{
\begin{tabular}{c c c c c c c c}
\hline
\hline
$a ["]$ & $e$ & $i[^{\circ}]$ & $\Omega[^{\circ}]$ & $\omega[^{\circ}]$ & $t_P$[yr]& T[yr]&K \\
\hline
0.123 $\pm$ 0.001 & 0.88 $\pm$ 0.003 & 135.25 $\pm$ 0.47 & 225.39 $\pm$ 0.84 & 63.56 $\pm$ 0.84 & 2002.32 $\pm$ 0.01& 15.8 $\pm$ 0.11& 14  \\
\hline
\end{tabular}}

\end{table*}
In the strong deflection limit, for very large bending angles ($\alpha > \pi$), the bending angle is of the form:
\begin{equation}
\alpha(\theta)=\overline{a} \log \left (\frac{\theta }{\theta_{\rm m}}-1 \right)+\overline{b},
\label{eq:bendingbozza2}
\end{equation}
where $\theta_{\rm m} \equiv u_{\rm m}/D_L$. The quantities $\overline{a}$ and $\overline{b}$ are functions of the metric determined for an arbitrary static spherically symmetric, metric in \cite{bozza2007}. The values of the functions for the three cases considered are in Table \ref{tab:ab}. The quantity $\theta_{\rm m}$ represents the angular size of the photon sphere to the observer. The bending angle is solved for $\theta$ and its dependence on time is
\begin{equation}
\theta(t) = \theta_{\rm m} \left [1+ e^{\left (\overline{b}-\gamma(t) \right)/\overline{a}}\right].
\label{eq:thetasdl}
\end{equation}
The magnification is derived from Eq. (\ref{eq:mag}). The definition of $\theta$ in Eq. (\ref{eq:thetasdl}) makes it straightforward to calculate $d \theta / d\gamma$. The magnification is then
\begin{equation}
\mu(t)= -\frac{D_S}{D_{LS}^2(t)} \frac{\theta_{\rm m}^2 e^{ \left (\overline{b}-\gamma(t) \right)/\overline{a}} \left[1+  e^{ \left (\overline{b}-\gamma(t) \right)/\overline{a}} \right]}{\overline{a} \sin \gamma(t)}.
\label{eq:magsdl}
\end{equation}
The formulations in Eqs. (\ref{eq:thetasdl}) and (\ref{eq:magsdl}) are particularly powerful because the time dependence of $D_{LS}$ and $\gamma$ are determined by defined orbital parameters.   

\begin{table}[b]
\caption{\label{tab:ab} $\overline{a}$ and $\overline{b}$ are functions of the metric that change for different values of $q$. In this table we list the values of these functions for the TRN ($q= -1.6$), ERN ($q = 0.25$) , and Schwarzschild cases ($q=0$).}
\begin{tabular}{c c c}
\hline
\hline
Case & $\overline{a}$& $\overline{b}$\\
\hline
TRN & $0.833$ & $-0.508$\\
Sch. & $1$ & $-0.400$\\
ERN & $1.414$ & $-0.733$ \\
\hline
\end{tabular}
\end{table}

To consider the case of a relativistic image, we consider the source to be at $\gamma(t) + 2 \pi$. This reflects the scenario of light looping around the black hole before reaching the observer. We consider this the first relativistic image, and we examine its properties in the next section.

\subsection{Relativistic Image Positions and Brightness}
\label{sec:reltheta}

\begin{figure}[h]
 \begin{center}
\includegraphics[width=0.7 \textwidth]{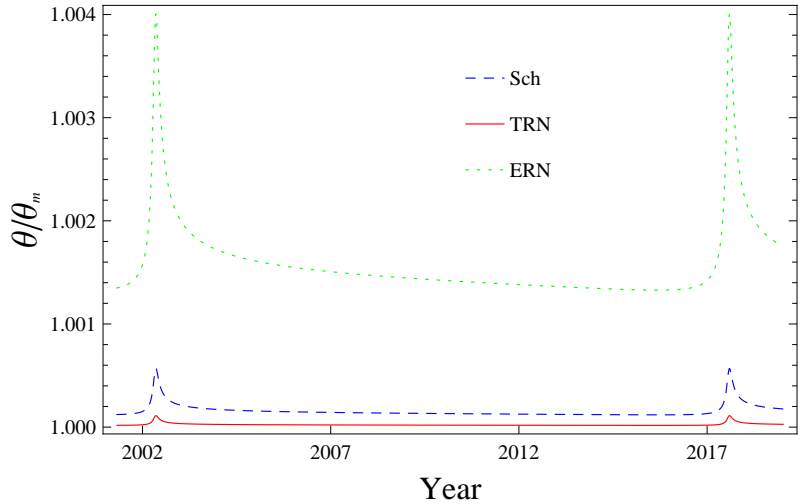}
 \caption{The ratio of $\theta/\theta_{\rm m}$ over time for S2for slightly more than one orbital period.}
 \label{fig:reltheta}
 \end{center}
 \end{figure}

Relativistic images form very close to the angular position of the photon sphere. This is because light from a star would need to pass close enough to the photon sphere to generate a large enough deflection for it to loop around the black hole before reaching the observer. The greater the bending angle required to reach the observer, the closer light must approach to the black hole. We would like to analyze how relativistic images appear relative to the photon sphere in the Schwarzschild, TRN ($q= -1.6$), and the ERN ($q= 0.25$) spacetimes. The size of the photon sphere is different in each spacetime, but we consider the image's position relative to the photon sphere. For an observer on Earth, the size of the photon sphere is  $ \theta_{\rm m}=27.5 \: \mu$ arc sec in a Schwarzschild spacetime, $\theta_{\rm m}=44.8 \: \mu$ arc sec in a TRN spacetime, and $\theta_{\rm m}=21.2 \: \mu$ arc sec in an ERN spacetime. We are more interested in the relationship between the image position and the photon sphere over time in each spacetime, which we demonstrate  in Fig. \ref{fig:reltheta}, than in the actual position of the image over time. In Fig. \ref{fig:reltheta}, we show the ratio of the image position to $\theta_{\rm m}$ over time.

There are several interesting results in Fig. \ref{fig:reltheta}. First is that although the image position in the ERN spacetime is smaller than the image position in the other two spacetimes, the image shifts considerably more relative to the photon sphere.  In terms of the different spacetimes, relativistic images appear closest to the photon sphere in the TRN spacetime, a little more distant from the photon sphere in the Schwarzschild spacetime, and the most shifted images appear in an ERN spacetime. Secondly, the shift relative to the photon sphere, for all spacetimes, in smallest for S2, greater for S14, and greatest for S6 (the plots are not shown for S6 and S14). Understanding these effects can give us insight into the strong lensing that takes place near a black hole.

\begin{figure}[h]
 \begin{center}
\includegraphics[width=0.7 \textwidth]{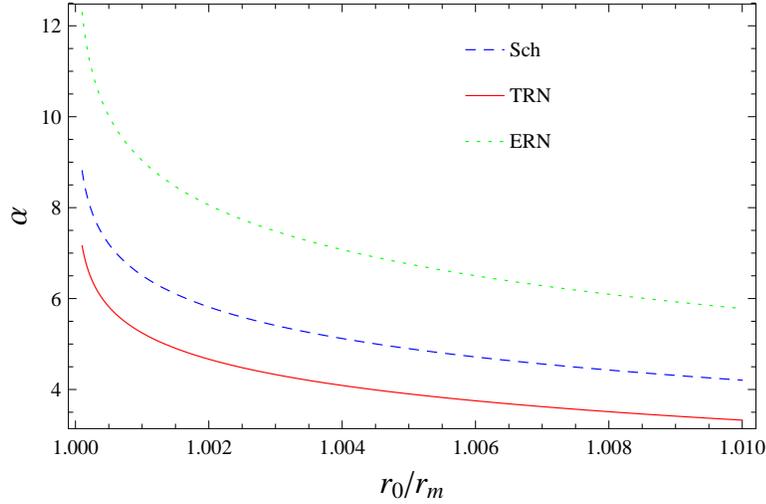}
 \caption{This figure displays $\alpha(r)$ for our usual three metrics. The point of closest approach is given in coordinates normalized by $r_{\rm m}$.  Here we show the bending angle very close to the photon sphere. In this way, we see the dependence of the bending angle on distance from the photon sphere. As can be seen, the bending angle for the ERN is noticeably larger than the other two near the photon sphere.}
 \label{fig:bendingspacetime}
 \end{center}
 \end{figure}

The images appear close to the photon sphere for S2 (when compared to S6 or S14). This is due to the poor alignment of S2 with the optic axis even at its peak alignment ($\gamma_0 = 45.3^{\circ}$). Light must pass close to the photon sphere to generate the extra deflection required to make up for the large source angular position. We can explain the relationship between the spacetime and position of the relativistic image relative to the photon sphere by considering the plot of the bending angle in each spacetime. In Fig. \ref{fig:bendingspacetime}, we plot the bending angle of the three spacetimes as a function of closest approach. The bending angle function is steepest for the ERN metric. As the point of closest approach approaches the photon sphere, the bending angle in the ERN spacetime is largest. The bending angle for the Schwarzschild spacetime does not descend as quickly, and the descent is gentlest for the TRN spacetime. For a given source, lens, and observer configuration, the bending angle is determined. By inspection of Fig. \ref{fig:bendingspacetime}, the value of $r_0/r_{\rm m}$ required to generate that bending angle is smallest for the TRN spacetime and largest for the ERN spacetime. Hence, the image in the ERN spacetime appears furthest from the photon sphere on a relative (to the photon sphere) basis.

\begin{figure}[h]
 \begin{center}
\includegraphics[width=0.7 \textwidth]{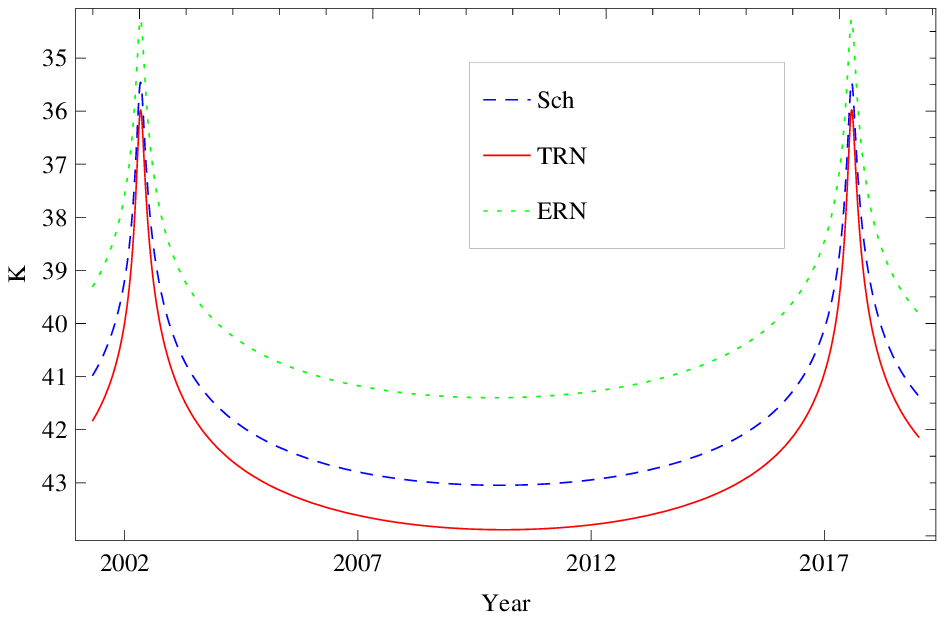}
\includegraphics[width=0.7 \textwidth]{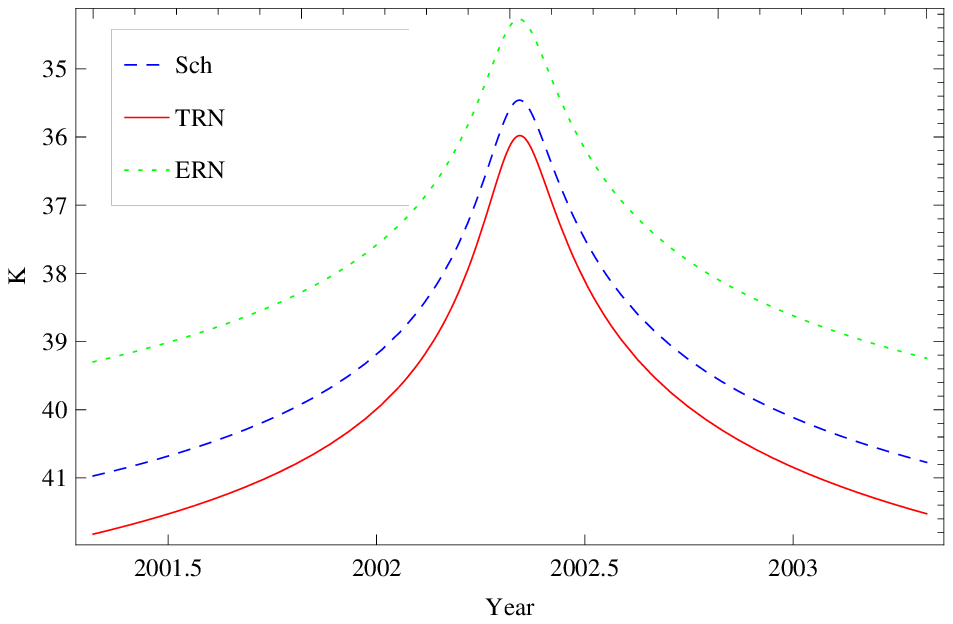} 
\caption{The brightness in the $K$ band of the first relativistic image of S2 in three different spacetimes over the course of a little more than an orbital period (top). The lower panel displays a close up of the brightness in the period surrounding peak brightness.}
 \label{fig:s2rel}
 \end{center}
 \end{figure}

Next, we used the formula for image magnification in Eq. (\ref{eq:magsdl}) to calculate the apparent brightness of the first relativistic image of S2 over the course of its orbit around Sgr A*. This is shown in Fig. \ref{fig:s2rel}.

\begin{figure}[h]
 \begin{center}
\includegraphics[width=0.7 \textwidth]{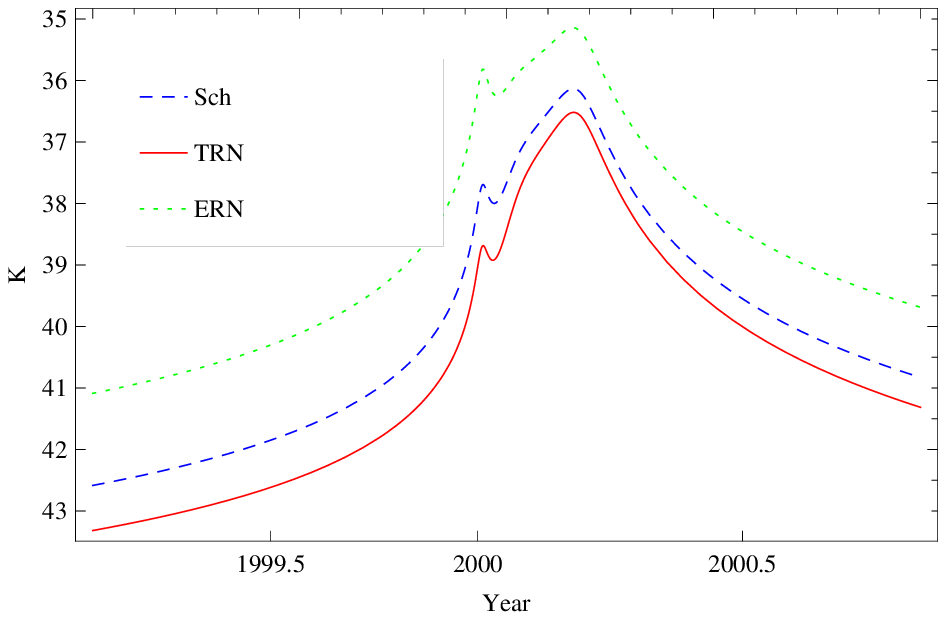}
\includegraphics[width=0.7 \textwidth]{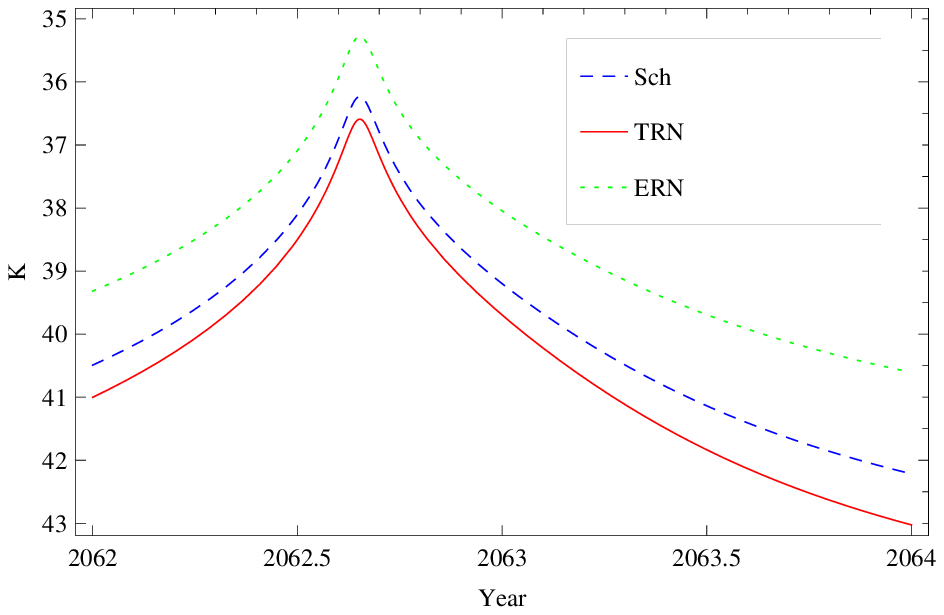} 
\caption{The brightness in the $K$ band of the first relativistic image of S14 in three different spacetimes over the course of about a year near its beak brightness (top). The brightness in the $K$ band of the first relativistic image of S6 in three different spacetimes over the course of about a year near its beak brightness (top)}
 \label{fig:s6rel}
 \end{center}
 \end{figure}

At peak brightness, and at all other times, the brightest image is the one in the ERN spacetime, which constrasts with the case of the secondary image, in which the TRN spacetime is the brightest. Even though image positions in the ERN spacetime are smaller and, therefore, tangential magnification is smaller, the radial magnification is increased in the ERN spacetime because of the shallower nature of $\alpha(r_0)$ in the ERN spacetime. At peak, the ERN image is about 1.2 magnitudes brighter than the Schwarzschild image and about 1.7 magnitudes brighter than the TRN image. When the images are fainter, the differences are even greater, as many as 2.5 magnitudes can separate the ERN image and the fainter TRN image, and the ERN image can be as many as 1.7 magnitudes brighter than the Schwarzschild image. The peak brightness of all images, both secondary and relativistic, occurs during the period between periapse and maximum alignment about 13 days later. Interestingly, the peak brightness of relativistic images occurs a bit earlier than the peak brightness of the secondary image. And even within relativistic images, peak brightness occurs at slightly different times for each metric. In terms of time after periapse, the ERN image peaks about $7.9$ days after periapse, the Schwarzschild image $8.4$ days, the TRN image $8.7$ days, and the secondary image $10.4$ days. As S2 moves away from periapse, the $1/D_{LS}^2$ term in the magnification contributes towards dimming the image. However, as $\gamma$ continues to grow smaller as S2 approaches its maximum alignment, the term $d \theta / d \gamma$ grows and contributes to brightening the image. For part of the period between periapse and maximum alignment, the increase in  $d \theta / d \gamma$ offsets the decrease in $1/D_{LS}^2$ and the image continues to brighten. However, as $\gamma$ grows smaller, the photons do not go as deeply into potential well as they require a smaller bending angle. Since the bending angle function in all spacetimes flattens as $r_0$ gets larger, $d \theta / d \gamma$ grows at a smaller rate and eventually, the decreasing $1/D_{LS}^2$ term dominates and the image starts to grow fainter. For an image in an ERN spacetime $r_0/r_{\rm m}$ is greater than it is for a TRN or Schwarzschild spacetime --- for a given configuration and required value of $\alpha$, light in the TRN requires a path that brings it closer to the photon sphere. In an ERN spacetime, however, a larger value of $r_0/r_{\rm m}$ generates the same bending angle. Therefore, the image in the ERN spacetime is in a flatter part of the bending angle function in Fig. \ref{fig:bendingspacetime}, and the bending angle does not decrease as quickly as $\gamma$ decreases. Hence, the radial magnification does not go up as quickly as $\gamma$ approaches periapse, and the image reaches peak brightness quicker. Since the image in the TRN spacetime is stuck deeper in the black hole's potential, near the photon sphere, as $\gamma$ decreases, the bending angle decreases quicker, leading to a greater increase in tangential magnification, and the image continues to grow brighter. A similar dynamic plays out for other stars we have analyzed, and the light curves for the first relativistic image of S6 and S14 are displayed in Fig. \ref{fig:s6rel}.

\section{Conclusion}
\label{sec:conc}
We have analyzed the difference between relativistic images in extremal Reissner-Nordstrom, tidal Reissner-Nordstrom, and Schwarzschild spacetimes. Although observational capabilities are nowhere close to being able to see these images, and the uncertainties in these stars' orbital parameters are more than some of the effects discussed,  the study of these details challenges and expands our understanding of the magnification mechanism, dynamics in the vicinity of a black hole, and the subtle effects that come into play for relativistic images. These lessons can be used for futher studies of images in the galactic center. The fact that the properties of images are sensitive to the choice of metric indicates that gravitational lensing in the galactic center may one day be an important probe of GR.


\section*{References}
\begin{thebibliography}{10}

\bibitem{LensingReview}
Wambsganss J 1998 Gravitational lensing in astronomy \newblock {\em Living Reviews in Relativity} 1 ; Schneider P  Ehlers J, and  Falco E E 1992 {\em Gravitational Lenses}. (Berlin: Springer); Hoekstra H and Jain B 2008 Weak Gravitational Lensing and its cosmological applications {\em Annual Review of Nuclear and Particle Science} 58 99


\bibitem{propSgrA}
Doelman S \etal 2008 Event-horizon-scale structure in the supermassive black hole  candidate at the galactic centre {\em Nature} 455; Ghez A M \etal 2008 Measuring distance and properties of the Milky Way's central supermassive black hole with stellar orbits {\em Astrophys. J.} 689 1044; Gillessen S \etal. 2009 \newblock {Monitoring stellar orbits around the massive black hole in the
 Galactic Center} \newblock {\em Astrophys. J.} 692 1075

\bibitem{VE2002}
Virbhadra K S and Ellis G F R 2002 \newblock Gravitational lensing by naked singularities{\em Phys. Rev. D}, 65 103004

\bibitem{darwin}
Darwin C 1959 The gravity field of a particle {\em Royal Society of London Proceedings Series A}, 249 180; Ohanian H C 1987 {\em Am. J. Phys.} 55 428; Misner C W, Thorne K S and Wheeler J A 1973 {\em {Gravitation}}

\bibitem{dmptidalrn}
Dadhich N, Maartens R, Papadopoulos P, and Rezania V 2000 \newblock {\em Phys. Lett. B} 487 1

\bibitem{whiskerthesis}
Whisker R 2008 \newblock {\em gr-qc/0810.1534}

\bibitem{me}
Bin-Nun A Y 2010 \newblock Relativistic images in Randall-Sundrum II braneworld lensing {\em Phys. Rev. D} 81 123011

\bibitem{me2}
Bin-Nun A Y 2010 \newblock Gravitational lensing of stars orbiting Sgr A* as a probe of the black hole metric in the galactic center. \newblock {\em Phys. Rev. D} 82 064009

\bibitem{methesis}
Bin-Nun A Y 2010 \newblock {\em Gravitational Lensing with a Large Bending Angle as a Probe of
  General Relativity and the Galactic Center} \newblock PhD thesis, University of Pennsylvania

\bibitem{bozzareview}
Bozza V 2010 \newblock {Gravitational lensing by black holes} {\em Gen. Rel. Grav.} 42 2269

\bibitem{VE2000}
{Virbhadra} K S and {Ellis} G F R 2000 \newblock {Schwarzschild black hole lensing}.\newblock {\em Phys. Rev. D}  62 084003

\bibitem{VK2008}
Virbhadra K S and Keeton C R 2008 Time delay and magnification centroid due to gravitational lensing by naked singularities.
\newblock {\em Phys. Rev. D}  77 124014

\bibitem{bozzaetal}
Bozza V, Capozziello S, Iovane G, and Scarpetto G 2001 \newblock {\em Gen. Rel. Grav.} 33 1535

\bibitem{bozza2002}
Bozza V 2002 \newblock {\em Phys. Rev. D} 66 103001

\bibitem{bozza2007}
{Bozza} V and {Scarpetta} G 2007 \newblock {Strong deflection limit of black hole gravitational lensing with
  arbitrary source distances}. \newblock {\em Phys. Rev. D} 76 083008

\bibitem{eiroad}
Eiroa E F 2005 \newblock Braneworld black hole gravitational lens: Strong field limit
  analysis \newblock {\em Phys. Rev. D} 71 083010

\bibitem{eiroarn}
Eiroa E F,  Romero G E, and Torres D F 2002 \newblock {Reissner-Nordstr{\"o}m black hole lensing} \newblock {\em Phys. Rev. D} 66 024010

\bibitem{whisker2005}
Whiser R 2005 \newblock {Strong gravitational lensing by braneworld black holes} \newblock {\em Phys. Rev. D} 71 064004

\bibitem{casadio}
Casadio R, Fabbri A, and Mazzacurati L 2002 \newblock {New black holes in the brane world?} \newblock {\em Phys. Rev. D} 65 084040

\bibitem{mmreview}
Majumdar A and Mukherjee N. 2005 \newblock Braneworld black holes in cosmology and astrophysics. \newblock {\em International Journal of Modern Physics D} 14 1095

\bibitem{aliev}
Aliev A N and Talazan P 2009 \newblock {Gravitational effects of rotating braneworld black holes} \newblock {\em Phys. Rev. D} 80 044023

\bibitem{exotic1}
Eiroa E F and Sendra C M 2010 Gravitational lensing by a regular black hole \newblock {\em ArXiv e-prints gr-qc/1011.2455}

\bibitem{exotic2}
Liu Y, Chen S, and Jing J 2010 Strong gravitational lensing in a squashed Kaluza-Klein black hole
  spacetime \newblock {\em Phys. Rev. D} 81 124017

\bibitem{exotic3}
Ghosh T and Sengupta S 2010 Strong gravitational lensing across a dilaton anti-de Sitter black hole \newblock {\em Phys. Rev. D} 81 044013

\bibitem{micado}
Trippe S \etal 2009 \newblock {\em Mon. Not. R. Astron. Soc.} 402 1126

\bibitem{holzwheeler}
Holz D E and Wheeler J A 2002 Retro-MACHOs: {$\pi$} in the Sky? \newblock {\em Astrophys. J.} 578 330

\bibitem{depaolis}
De Paolis F, Geralico A, Ingrosso G, and Nucita A A 2003 The black hole at the galactic center as a possible retro-lens for
  the S2 orbiting star {\em Astron. Astrophys.} 409 809

\bibitem{bozzas2}
Bozza V and Mancini L 2004 \newblock {\em Astrophys. J.} 611 1045; Bozza V and Mancini L 2005 \newblock {\em Astrophys. J.} 627 790; Bozza V and Mancini L 2009 \newblock {\em Astrophys. J.}, 696 435

\bibitem{bozzalens}
Bozza V 2008 Comparison of approximate gravitational lens equations and a proposal for an improved new one \newblock {\em Phys. Rev. D} 78 103005

\end{thebibliography}
\end{document}